\documentclass[11pt]{article}
\usepackage{a4wide}
\usepackage{epsfig,epsf,graphics,psfrag}
\usepackage{times}
\usepackage{float}
\usepackage{color}
\usepackage{amsfonts,amssymb,stmaryrd,latexsym,amsmath}
\usepackage{multirow}
\usepackage{array} 

\newcommand{\Ups}{\Upsilon(3S)\to\Upsilon(2S)\pi^0\pi^0}

\begin{document}
\title{Extracting $\pi\pi$ $S$-wave scattering lengths from cusp effect in heavy quarkonium dipion transitions}

\author{Xiao-Hai Liu$^1$\footnote{{\it E-mail address:}
Xiao-Hai.Liu@tp2.ruhr-uni-bochum.de},  \
Feng-Kun~Guo$^2$\footnote{{\it E-mail address:}
fkguo@hiskp.uni-bonn.de}, \
Evgeny Epelbaum$^1$\footnote{{\it E-mail address:} evgeny.epelbaum@rub.de}
     \\
     {\it\small$\rm ^1$ Institut f\"ur Theoretische Physik II, Rhur-Universit\"at Bochum, D-44780 Bochum, Germany
            } \\
    {\it\small$\rm ^2$Helmholtz-Institut f\"ur Strahlen- und Kernphysik and
          Bethe Center for Theoretical Physics,}\\
          {\it\small Universit\"at Bonn, D--53115 Bonn, Germany}
          }

\maketitle

\begin{abstract}

Charge-exchange rescattering $\pi^+\pi^-\to \pi^0\pi^0$ leads to a cusp
effect in the $\pi^0\pi^0$ invariant mass spectrum of processes with $\pi^0\pi^0$
in the final state which can be used to measure $\pi\pi$ $S$-wave
scattering lengths. Employing a non-relativistic effective field theory, we discuss the
possibility of extracting the scattering lengths in heavy quarkonium $\pi^0\pi^0$
transitions. The transition $\Upsilon(3S)\to\Upsilon(2S)\pi^0\pi^0$ is studied in
details. We discuss the precision that can be reached
in such an extraction for a certain number of events.

\noindent
\end{abstract}



\thispagestyle{empty}
\newpage


\section{Introduction}

Being much lighter than all the other hadrons, the pions play a
unique role in the strong interacti\-ons. They are the
pseudo-Goldstone bosons of the spontaneous chiral symmetry breaking
in quantum chromodynamics (QCD). Thus, to a large extent, the
interaction between pions are governed by spontaneous and explicit
chiral symmetry breaking. The $\pi\pi$ scattering problem already
has a long history, which began about half a century
ago~\cite{Weinberg:1966kf}. At low energies, the strength of the
$\pi\pi$ $S$-wave interaction is described by the scattering
lengths, which can shed light on the fundamental properties of QCD.
The scattering lengths can be calculated in chiral perturbation
theory (ChPT)~\cite{Weinberg:1978kz,Gasser:1983yg}, the low-energy
effective theory of QCD, to a given order in the chiral expansion.
Combining two-loop ChPT with Roy equations, the $\pi\pi$ scattering
lengths were predicted with a high
precision~\cite{Colangelo:2000jc,Colangelo:2001df}. For instance,
the difference between the isospin $I=0$ and $I=2$  $S$-wave
scattering lengths was predicted to be $(a_0-a_2)M_{\pi^+}=0.265\pm
0.004$. A similar result of $0.262\pm0.006$ was obtained in
Ref.~\cite{GarciaMartin:2011cn} using dispersion relations without
input from ChPT.

Experimentally, the $\pi\pi$ scattering lengths can be measured in several ways.
The angular distributions of $K_{e4}$ decay is sensitive to the $\pi\pi$ phase
shifts which are related with the scattering lengths. The first
experiment along these lines was carried out by the Geneva-Saclay
Collaboration in the seventies of the lest century
\cite{Rosselet:1976pu}.  A similar method was recently
employed  by the E865 and NA48/2 Collaborations
\cite{Pislak:2001bf,Batley:2007zz,Ananthanarayan:2000ht,Pislak:2003sv}.
Pionium lifetime can also be related to the $\pi\pi$
scattering lengths, and the experimental result is well consistent with the
prediction~\cite{Adeva:2005pg}. Another precise method relies on measuring the cusp
effect in the decay $K^+\to \pi^0\pi^0\pi^+$, which results from the
charge-exchange rescattering $\pi^+\pi^-\to
\pi^0\pi^0$~\cite{budini:1961,Meissner:1997fa,Cabibbo:2004gq,Cabibbo:2005ez,Colangelo:2006va,Batley:2005ax,Batley:2000zz,Gasser:2011ju}.
The cusp structure was also discussed in other processes including the $K_L\to
3\pi$, $\eta\to 3\pi$ and $\eta'\to
\eta\pi\pi$~\cite{Bissegger:2007yq,Ditsche:2008cq,Kubis:2009sb}.
Since the branching ratio of $K^+\to \pi^+\pi^-\pi^+$,  $(5.59\pm 0.04)\%$, is much
larger than that of $K^+\to \pi^0\pi^0\pi^+$, $(1.761\pm 0.022)\%$~\cite{PDG}, the
charge-exchange rescattering turns out to be important so that the
cusp effect in the $\pi^0\pi^0$ invariant mass spectrum appears to be
enhanced. However, it is difficult to accurately measure the cusp effect in $K_L\to 3\pi$ and $\eta\to 3\pi$ according to
currently available experimental data. The process $\eta'\to \eta\pi\pi$ is a promising candidate, in
which the cusp effect is predicted to have an effect of more than $8\%$ in the
decay spectrum below $\pi^+\pi^-$ threshold~\cite{Kubis:2009sb}. For a brief review
of the $\pi\pi$ scattering and a list of experimental measurements see
Ref.~\cite{Gasser:2009zz}.

In this paper, we investigate the possibility of extracting the $\pi\pi$ scattering
lengths using the cusp effect in heavy quarkonium dipion transitions. These
transitions are among the most important decay modes of the heavy quarkonium states
below open heavy-flavor thresholds.  Taking the process $\psi'\to J/\psi
\pi^0\pi^0$ as an example, the branching fraction is $17.75\pm0.34\%$~\cite{PDG},
and the BESIII and CLEO-c Collaborations have already accumulated huge data samples
in this channel. In particular, the BESIII Collaboration has acquired a sample of
106 million $\psi'$ events, and this number is still increasing \cite{Li:2011bi}.
Because of the Watson final-state theorem~\cite{Watson:1954uc}, it is possible to
learn about the $\pi\pi$ interaction from the dipion transitions. There were
suggestions of studying the $\pi\pi$ scattering phase shifts in the $\psi'\to
J/\psi\pi^+\pi^-$ transitions~\cite{Guo:2006fv,Zhang:2008tm,Chen:2009zzr}. There
are also huge data samples for the bottomonium states which were collected in the
$B$-factories, and more are expected to come from the next-generation
high-luminosity $B$-factories~\cite{Bevan:2012hi,Biagini:2012zz}. In view of this
situation, it is interesting to explore the cusp effect in heavy quarkonium
transitions with two neutral pions in the final state. Because the isospin symmetry
is well conserved here, one has $\mathcal{B}(\psi'\to J/\psi
\pi^+\pi^-)/\mathcal{B}(\psi'\to J/\psi \pi^0\pi^0)\approx 2$~\cite{PDG}, and
similarly for the other heavy quarkonium dipion transitions. This is similar to the
$\eta'\to \eta\pi\pi$, which will make the charge-exchange rescattering effect more
important than those in the processes $K_L\to 3\pi$ and $\eta\to 3\pi$. In
addition, it was found in Ref.~\cite{Kubis:2009sb} that two-loop rescattering is
highly suppressed in the $\eta'\to\eta\pi\pi$ process due to the approximate
isospin symmetry, so that the cusp in the $\pi^0\pi^0$ distribution is completely
dominated by one-loop contributions. The same conclusion should hold in our case.
Therefore, it is safe to work up to only one-loop order. From the theoretical point
of view, since the interaction between a heavy quarkonium and pion is highly
Okubo-Zweig-Iizuka (OZI) suppressed, we may further simplify the problem by
neglecting this type of contributions.

If we concentrate on the region near the $\pi\pi$ threshold, the
three-momenta of all the final particles are small in comparison
with their masses. Thus, a nonrelativistic effective field theory
(NREFT) can be employed. The processes to be considered are
similar to the $\eta'\to \eta\pi\pi$. We will follow here the NREFT
framework developed and applied in
Refs.~\cite{Gasser:2011ju,Kubis:2009sb} and refer to these papers and
references therein for more details. This method was firstly used in the study of
cusp effect in $K\to 3\pi$, and then extended to other reactions such
as e.g.~$\eta'\to \eta\pi\pi$ and $\eta\to 3\pi$.

Our paper is organized as follows. The framework of NREFT will be
briefly introduced in Section~\ref{sec:NREFT}, where the necessary
terms in the effective Lagrangians are listed. The low-energy
constants entering the tree-level
production amplitude are determined  in Section~\ref{sec:MC} by matching to a
relativistic description of the decay for the transition $\Ups$ in the
framework of unitarized chiral perturbation theory.
In the same section, we
generate several sets of synthetic data using the Monte Carlo
method, and investigate the accuracy of the extraction of the $\pi\pi$
scattering lengths from these synthetic data. A short summary is
presented in Section~\ref{sec:sum}, and an estimate of the
$J/\psi\pi$ scattering length is attempted in
Appendix~\ref{sec:app}.

\section{Nonrelativistic effective field theory}
\label{sec:NREFT}

Here and in what follows, we consider  $\psi'$ dipion transitions as an example.
The method can be easily extended to bottomonium case. Let us describe the power
counting scheme in the NREFT, which is essentially a nonrelativistic velocity
counting. The masses of all involved particles are counted as $\mathcal{O}(1)$.
Since we focus on the region close to the thresholds of two pions, even pions can
be dealt with nonrelativistically. The heavy quarkonium in the final state is also
nonrelativistic because its three-momentum in the rest frame of the decaying
particle does not exceed 500~MeV. We, therefore, count all these three-momenta as
quantities of order $\mathcal{O}(\epsilon)$. The kinetic energy $T_i=p_i^0-M_i$ is
then counted as $\mathcal{O}(\epsilon^2)$. Another expansion parameter used in this
scheme is the $\pi\pi$ scattering length, denoted by $a_{\pi\pi}$. Here one relies
on the fact that low-energy interactions between two pions are weak due to their
Goldstone boson nature. In principle, $J/\psi \pi$ scattering should also be taken
into account. However, it should be suppressed according to the OZI rule because
the $J/\psi$ and pion do not have any common valence quark. A rough estimation of
the $J/\psi \pi$ scattering length carried out in the appendix yields
$\left|a_{J/\psi\pi} \right|\lesssim0.02$~fm which is consistent with the
preliminary lattice result $a_{J/\psi\pi}=(-0.01 \pm 0.01)
\mbox{fm}$~\cite{Liu:2008rza}. Thus, the $J/\psi\pi$ scattering length is at least
one order of magnitude smaller than $a_0-a_2$. The bottomonium-pion scattering
length would be even smaller. The situation here is, therefore, similar to the one
in the process $\eta'\to \eta\pi\pi$, where $\eta\pi$ interaction was found to play
a minor role in the $\pi \pi$ cusp structure, and its effect can be largely absorbed into
the polynomial production amplitude~\cite{Kubis:2009sb}. Thus, in the
following, we will not take into account the heavy quarkonium-pion scattering.

The relevant effective Lagrangians contain two parts
\begin{eqnarray}
\mathcal{L}_{eff}=\mathcal{L}_\psi +\mathcal{L}_{\pi\pi}.
\end{eqnarray}
Here, the first term describes the production mechanism and reads up to
$\mathcal{O}(\epsilon^2)$
\begin{eqnarray}
\mathcal{L}_\psi &=& \frac{1}{2}\sum\limits_{n=0}^{1} G_n \left(
{\psi'}_i^{\dagger} (W_{J/\psi}-M_{J/\psi})^n J_i \Phi_0\Phi_0 +
h.c. \right) \nonumber \\
&+&  \sum\limits_{n=0}^{1} H_n \left( {\psi'}_i^{\dagger}
(W_{J/\psi}-M_{J/\psi})^n J_i \Phi_+\Phi_- + h.c. \right) +\cdots,
\label{eq:Lpsi}
\end{eqnarray}
where $W_{J/\psi}=\sqrt{M_{J/\psi}^2-\triangle}$, with $\triangle$ being the
Laplacian. At this order, the production is purely $S$-wave, while the $D$-wave
contribution starts from $\mathcal{O}(\epsilon^4)$. $\pi\pi$ interaction is
described by~\cite{Bissegger:2007yq}:
\begin{eqnarray}
\mathcal{L}_{\pi\pi}&=&2\sum\limits_{k=0,\pm}\Phi^\dagger_k
W_k(i\partial_t-W_k)\Phi_k \nonumber \\
&+& C_x(\Phi_0^\dagger\Phi_0^\dagger \Phi_+\Phi_-
+h.c.)+\frac{1}{4}C_{00}(\Phi_0^\dagger\Phi_0^\dagger \Phi_0\Phi_0
+h.c.) \nonumber \\
&+&  D_x\left[ (\Phi_0^\dagger)_\mu (\Phi_0^\dagger)^\mu
\Phi_+\Phi_- +\Phi_0^\dagger \Phi_0^\dagger(\Phi_+^\dagger)_\mu
(\Phi_-^\dagger)^\mu +h.c.\right] \nonumber
\\
&+&\frac{1}{4}D_{00}\left[ (\Phi_0^\dagger)_\mu (\Phi_0^\dagger)^\mu
\Phi_0\Phi_0 +\Phi_0^\dagger \Phi_0^\dagger(\Phi_0^\dagger)_\mu
(\Phi_0^\dagger)^\mu +h.c.\right]+\cdots,
\end{eqnarray}
where
\begin{eqnarray}
(\Phi_k)_\mu&=&(\mathcal{P}_k)_\mu \Phi_k,\
(\mathcal{P}_k)_\mu=(W_k,-i\nabla) \nonumber \\
(\Phi_k^\dagger)_\mu &=& (\mathcal{P}_k^\dagger)_\mu
\Phi_k^\dagger,\ (\mathcal{P}_k^\dagger)_\mu=(W_k,i\nabla)
\end{eqnarray}
and $W_k=\sqrt{M_k^2-\triangle}$.
In Eq.~(3), the couplings $C_x$, $C_{00}$, $D_x$ and $D_{00}$ can
be obtained by matching the NREFT amplitude to the effective range expansion of
$\pi\pi$ scattering amplitudes~\cite{Ananthanarayan:2000ht}
\begin{eqnarray}
&&T^I(s,t) = 32\pi \sum\limits_{l=0}^{\infty}(2l+1)t^I_l(s)P_l(z), \nonumber \\
 &&\text{Re}\ t_l^I(s) = q_{ab}^{2l}\left[a^I_l+b^I_l
q_{ab}^2+\mathcal{O}(q_{ab}^4)\right],
\end{eqnarray}
where $t^I_l$ is the partial wave amplitude with angular momentum $l$ and isospin
$I$, $P_l(z)$ are the Legendre polynomials with $z=\cos{\theta}$, where $\theta$ is
the scattering angle in the center-of-mass system, and
$q_{ab}=\left[\lambda(s,M_a^2,M_b^2)/s\right]^{1/2}/2$ is the center-of-mass
momentum with $\lambda(a,b,c)=a^2+b^2+c^2-2(ab+bc+ac)$ being the K\"all\'en function. We
thus have the following relations,
\begin{eqnarray}
C_x &=& \frac{16\pi}{3}M_{\pi^+}(a_0-a_2)\left(1+\frac{\xi}{3} \right),
\quad
C_{00} =  \frac{16\pi}{3}M_{\pi^+}(a_0 + 2a_2)(1-\xi),
\nonumber \\
 D_x &=& \frac{4\pi}{3}M_{\pi^+}(b_2-b_0),
\quad
D_{00} =  \frac{4\pi}{3}M_{\pi^+}(b_0 + 2b_2),
\end{eqnarray}
where $\xi = \left(M_{\pi^+}^2-M_{\pi^0}^2\right)/M_{\pi^+}^2$, and the isospin
breaking in the $S$-wave scattering lengths has been considered at leading order in
ChPT \cite{Knecht:1997jw}. We have used the phase convention such that
$|\pi^+\rangle = - |1,+1\rangle$. Because we only consider $S$-wave
scattering here, we have denoted the $I=0$ and $I=2$ scattering lengths by $a_0$
and $a_2$ for brevity in the above equations.

\begin{figure}[t]
\includegraphics[width=0.7\hsize]{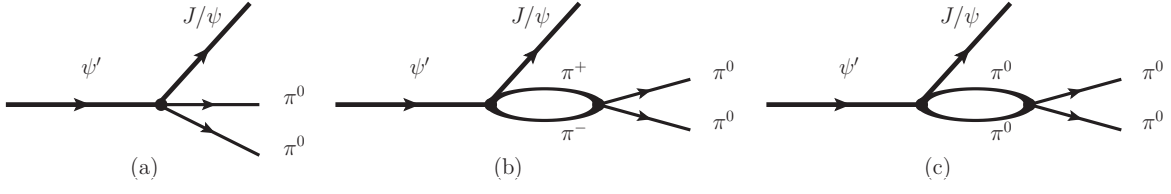}
\caption{$\psi'\to J/\psi\pi^0\pi^0$ via tree diagram and $\pi\pi$
rescattering diagrams.}\label{oneloop}
\end{figure}

In this paper, we  work only up to one-loop order. Neglecting the $J/\psi\pi$
interaction as explained above, the diagrams need to be considered are shown in
Fig.~\ref{oneloop}.
With the momenta defined as
\begin{eqnarray}
\psi'(P_{\psi'})\to \pi^0(p_1)\pi^0(p_2) J/\psi(p_3),
\end{eqnarray}
and $s_i=(P_{\psi'}-p_i)^2$ for $i=1, 2, 3$, the  transition amplitudes at the tree
and one-loop level are
\begin{eqnarray}
T^\text{\,tree}&=& \left[ G_0+G_1(p_3^0-M_J)  \right] \vec{\epsilon}_{\psi'} \cdot \vec{\epsilon}_{J}, \\
T^\text{\,1-loop} &=& 2 \left[C_x+D_x(s_3-4M_{\pi^+}^2) \right] \left[
H_0+H_1(p_3^0-M_J) \right] J_{+-}(s_3)\vec{\epsilon}_{\psi'} \cdot
\vec{\epsilon}_{J}
\nonumber\\
&+&  \left[C_{00}+D_{00}(s_3-4M_{\pi^0}^2) \right]\left[
G_0+G_1(p_3^0-M_J) \right] J_{00}(s_3)\vec{\epsilon}_{\psi'} \cdot
\vec{\epsilon}_{J},
\end{eqnarray}
respectively, where $J_{ab}$ is a nonrelativistic loop integral defined as
\begin{eqnarray}
J_{ab}(P^2)=\int \frac{d^D l}{i(2\pi)^D } \frac{1} {2w_a(\vec{l})
(w_a(\vec{l})-l_0) }  \frac{1} {2w_b(\vec{P}-\vec{l})
(w_b(\vec{P}-\vec{l})-P_0+l_0) },
\end{eqnarray}
with $w(\vec{l})=\sqrt{M^2+\vec{l}^2}$. Within the nonrelativistic power counting
scheme, the loop integral measure is counted as $\mathcal{O}(\epsilon^5)$, and each
of the two propagators is of order $\mathcal{O}(\epsilon^{-2})$. Thus, the loop
integrals $J_{+-}$ and $J_{00}$ are of order $\mathcal{O}(\epsilon)$. Using
dimensional regularization and taking $D=4$, we obtain
\begin{eqnarray}
J_{+-}(s_3) &=& \frac{-1}{16\pi}
\sqrt{\frac{4M_{\pi^+}^2-s_3}{s_3}},\ \ \mbox{when}\ s_3\leq
4M_{\pi^+}^2, \\
J_{+-}(s_3) &=& \frac{i}{16\pi}
\sqrt{\frac{s_3-4M_{\pi^+}^2}{s_3}},\ \ \mbox{when}\ s_3>
4M_{\pi^+}^2,\\
J_{00}(s_3) &=& \frac{i}{16\pi} \sqrt{\frac{s_3-4M_{\pi^0}^2}{s_3}}.
\end{eqnarray}
One observes that $J_{+-}$ has a nonanalyticity at the $\pi^+\pi^-$ threshold
which gives rise to a cusp effect in the $\pi^0\pi^0$ invariant mass
distribution. The expression for the decay amplitude up to
$\mathcal{O}(a_{\pi\pi}\epsilon^2)$ reads:
\begin{equation}
    T =  \left[ G_0+G_1(p_3^0-M_J)  + 2 C_x H_0 J_{+-}(s_3) +  C_{00}
 G_0 J_{00}(s_3) \right] \vec{\epsilon}_{\psi'} \cdot
\vec{\epsilon}_{J}.
\label{eq:NRT}
\end{equation}
In the isospin limit, we have $H_0=G_0$ in our phase convention, and it will be
used in the following.

\section{Extraction of the scattering lengths}
\label{sec:MC}

In this section we explore the possibility to extract the
$\pi\pi$ scattering lengths from the heavy quarkonium dipion transitions. A known
feature of the reaction $\psi'\to J/\psi\pi\pi$ is that the
kinematical region around the the $\pi\pi$
threshold we are interested in is strongly suppressed so that it only
corresponds to a
tiny fraction of the total events, see e.g. the BES and CLEO data for the $\psi'\to
J/\psi\pi^+\pi^-$~\cite{Bai:1999mj,Mendez:2008kb}. A more promising
reaction is the $\Ups$, which we will concentrate on. The updated data came from the
CLEO Collaboration~\cite{CroninHennessy:2007zz}, and their analysis is based on a
$\Upsilon(3S)$ yield about $5\times10^6$.

\subsection{Chiral unitary approach}
In order to show the cusp effect in the $\Ups$, it is necessary to determine the
values of $G_0$ and $G_1$. This will be achieved via matching to
parameters entering the relativistic
decay amplitude which can be fixed from fitting to the experimental data of the
$\pi^0\pi^0$ invariant mass spectrum. A precise determination is
presently not possible due to the bad data quality. However, a rough estimate of the ratio
$G_0/G_1$ can be obtained. We employ a simple parametrization of the tree-level
relativistic decay amplitude~\cite{Voloshin:1987rp}
\begin{equation}
    N \epsilon\cdot\epsilon' (s_3 - C),
    \label{eq:propoly}
\end{equation}
where $N$ is an overall normalization constant, and $\epsilon$
($\epsilon'$) is the polarization vector of $\Upsilon(2S)$
($\Upsilon(3S)$).
\begin{figure}[tb]
\centering
\includegraphics[width=0.6\hsize]{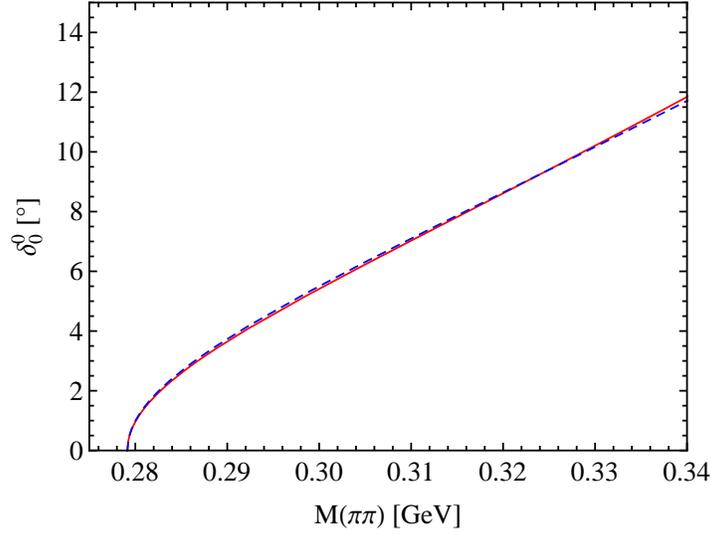}
\caption{Fit to the parametrization of $\pi\pi$ phase shifts introduced in Ref.~\cite{GarciaMartin:2011cn}
(solid line), where the dashed line represents the CHUA results.}\label{fig:ps}
\end{figure}
The cusp effect shows up only when the $\pi\pi$ FSI is considered. This may be
taken into account using the chiral unitary approach
(CHUA)~\cite{Kaiser:1995cy,Oller:1997ti,Oller:1998hw,Oller:1998zr,Oller:2000fj},
which has been used in studying the dipion transitions among heavy quarkonium
states in Refs.~\cite{Uehara:2002wh,Guo:2004dt,Guo:2006ya}. In the CHUA, the $\pi\pi$ $S$-wave
scattering amplitude after taking into account isospin symmetry is given by
\begin{equation}
T_0^0(s_3) = V_0^0(s_3)\left[1-G(s_3)V_0^0(s_3)\right]^{-1},
\label{eq:Tchua}
\end{equation}
where the $2\times2$ matrix $V_0^0(s_3)$ contains the $S$-wave projected
$\pi^0\pi^0\to\pi^0\pi^0$, $\pi^+\pi^-\to\pi^+\pi^-$ and $\pi^0\pi^0\to\pi^+\pi^-$
amplitudes derived from the lowest order chiral perturbation theory with virtual
photons~\cite{Ecker:1988te},
\begin{equation}
    V_0^0(s_3) = \frac1{F_\pi^2}\left(
    \begin{array}{cc}
        M_{\pi^0}^2/2 &  \left( s_3 - M_{\pi^0}^2 \right)/\sqrt{2} \\
        \left( s_3 - M_{\pi^0}^2 \right)/\sqrt{2} &
        \left[s_3 +4 \left(M_{\pi^+}^2-M_{\pi^0}^2 \right) \right]/2
    \end{array}
     \right).
\end{equation}
Here $F_\pi$ is the pion decay constant
and the difference 
between the charged and neutral pion masses
is taken into account.
$G(s_3)=\text{diag}\left\{ G_{00}(s_3), G_{+-}(s_3)\right\}$ is a
diagonal matrix with
\begin{equation}
    G_{00(+-)}(s_3) = -\frac1{16\pi^2} \left[ \tilde{a}(\mu) + \log \frac{M_{\pi^{0(+)}}^2}{\mu^2}
    + \sigma_{0(+)} \log \left( \frac{\sigma_{0(+)}+1}{\sigma_{0(+)}-1} \right) \right]
\end{equation}
denoting the usual scalar loop function. Here, $\sigma_{0(+)}=\sqrt{ 1 - 4
M_{\pi^{0(+)}}^2 /s_3 }$, and $\tilde{a}(\mu)$ is a subtraction constant introduced
to regularize the loop~\cite{Oller:1998zr,Oller:2000fj}. In order for the $\pi\pi$
FSI to be consistent with $\pi\pi$ scattering, the value of $\tilde{a}(\mu)$
may be fixed by reproducing the $S$-wave $\pi\pi$ phase shifts in the isoscalar
channel, $\delta_0^0(s_3)$. We fit to the parametrization of $\delta_0^0(s_3)$
introduced in Ref.~\cite{GarciaMartin:2011cn}, which is given by Eq.~(6) in that
paper, and the central values of the parameters $B_i$ in the parametrization are
used. Isospin breaking effects are neglected in the fit. The fit range is chosen to
be from the $\pi\pi$ threshold up to 340~MeV, which contains all the available
phase space of the $\Ups$. Our best fit is shown in Fig.~\ref{fig:ps}
and yields $\tilde{a}(1~\text{GeV}) = -0.930$.
\begin{figure}[t]
\centering
\includegraphics[width=\hsize]{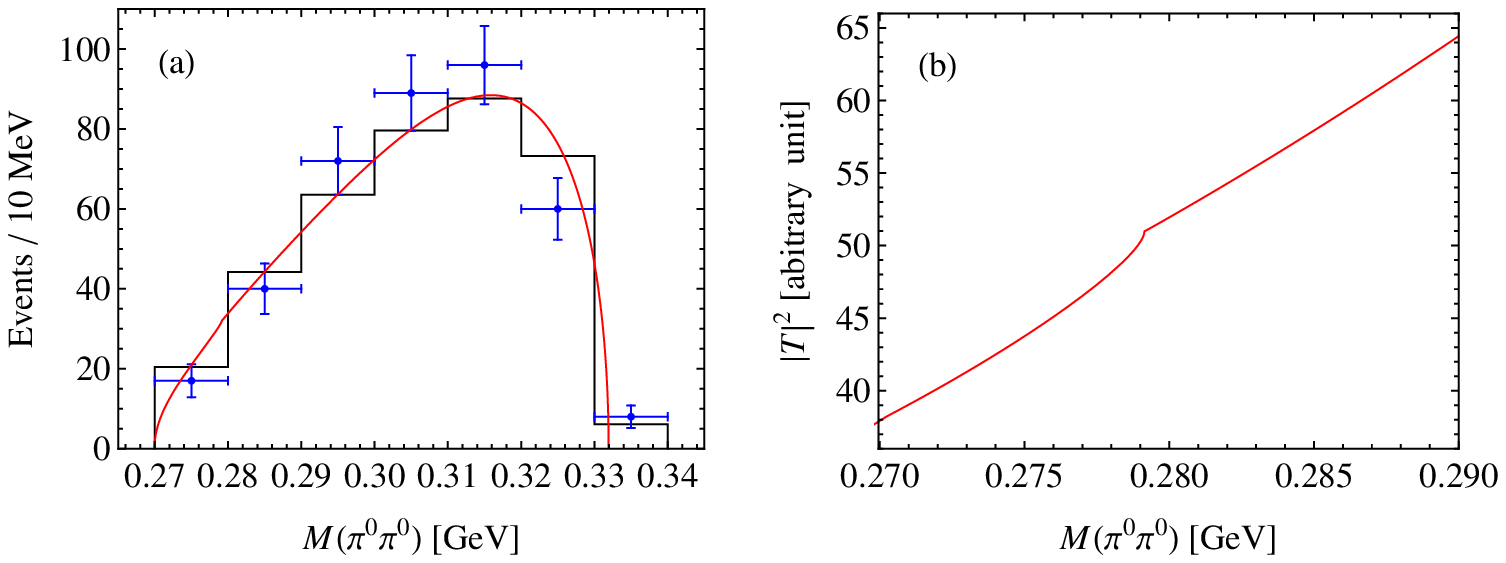}
\caption{(a) Comparison of the best fit (histogram) to the $\pi^0\pi^0$ invariant mass
spectrum of the $\Ups$ data (points with error bars) measured in
Ref.~\cite{CroninHennessy:2007zz}. The fit is done by integrating the distribution bin-by-bin.
The solid smooth curve is the invariant mass spectrum calculated using the best fit parameters
and multiplied by an arbitrary normalization constant.
(b) The phase space subtracted spectrum around the $\pi^+\pi^-$ threshold. }\label{fig:pipi}
\end{figure}
With this value of $\tilde{a}(1~\text{GeV})$, we can fix the value of $C$ in
Eq.~\eqref{eq:propoly} by fitting to the $\pi^0\pi^0$ invariant mass spectrum of
the $\Ups$ data measured in Ref.~\cite{CroninHennessy:2007zz}. The decay amplitude
in the CHUA is given by
\begin{equation}
    N \epsilon\cdot\epsilon' (s_3 - C) \left[ 1 + G_{00}(s_3) T_0^0(s_3)_{11} +
    \sqrt{2} G_{+-}(s_3) T_0^0(s_3)_{21}  \right],
\end{equation}
where $T_0^0(s_3)_{11(21)}$ refer to the unitarized amplitudes for the
$\pi^0\pi^0\to\pi^0\pi^0\, (\pi^+\pi^-\to\pi^0\pi^0)$ defined in
Eq.~\eqref{eq:Tchua}. The best fit with $\chi^2/dof=1.44$ is shown in
Fig.~\ref{fig:pipi}~(a). A small cusp at the $\pi^+\pi^-$ threshold shows up, which
is more apparent in the phase-space-subtracted invariant mass spectrum in
Fig.~\ref{fig:pipi}~(b). From the fit, we obtain
$$C =
-0.0197^{+0.0167}_{-0.0116}~\text{GeV}^2 = -1.01^{+0.86}_{-0.50}~M_{\pi^+}^2.$$
\begin{figure}[tbh]
\centering
\includegraphics[width=0.6\hsize]{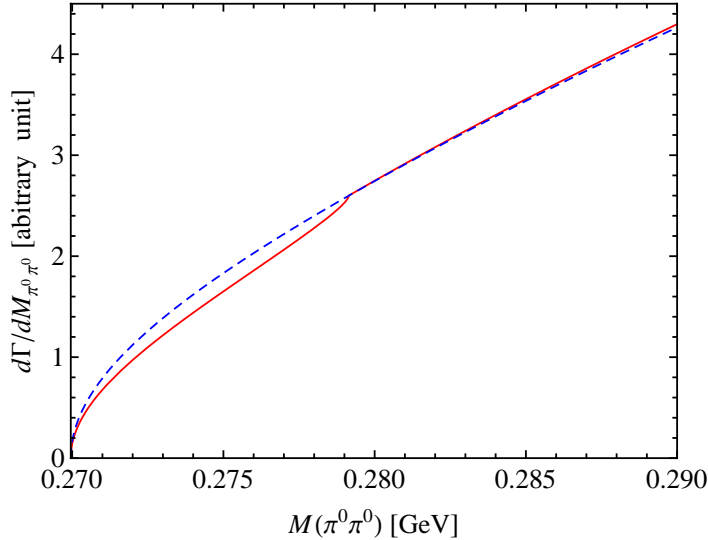}
\caption{The cusp effect at the $\pi^+\pi^-$ threshold in the reaction
  $\Ups$ calculated
  in the NREFT framework (solid line).
The dashed line shows the result without charge-exchange rescattering. }\label{fig:cusp}
\end{figure}
Matching $G_{0,1}$ to $N$ and $C$ in Eq.~\eqref{eq:propoly} leads to
the relations
\begin{equation}
    G_0 = N \left[ (M-m_3)^2 - C \right], \qquad G_1 = - 2 N M.
    \label{eq:G01}
\end{equation}
Thus, the ratio is determined to be
\begin{equation}
    \frac{G_0}{G_1} = -4.37^{+0.81}_{-0.56}~\text{MeV}.
    \label{eq:G01ratio}
\end{equation}
It is small because $G_0$ contains the mass difference of the two heavy quarkonia
while $G_1$ is proportional to the mass of the initial state. Using the central
value and adopting the central values of the scattering lengths summarized in
Ref.~\cite{Gasser:2009zz}, $a_0=0.2196$ and $a_2=-0.0444$ in units of
$M_{\pi^+}^{-1}$, the cusp effect in the NREFT is plotted in Fig.~\ref{fig:cusp}.
Certainly, if charge-exchange rescattering is switched off, the cusp would
disappear as shown by the dashed line. When integrating the spectrum below the
threshold of $\pi^+\pi^-$, the cusp effect resulted from charge-exchange
rescattering will reduce the number of events in this region by about $9\%$ with
respect to the tree-level contribution. The values of this quantity in the
processes $K^+\to \pi^0\pi^0\pi^+$, $\eta'\to \eta \pi\pi$ and $\eta\to 3\pi$ are
about $13\%$, $8\%$ and less than $2\%$,
respectively~\cite{Kubis:2009sb,Gullstrom:2008sy}.

\subsection{Monte Carlo simulations}

An important question is to what precision the scattering lengths
can be extracted from the considered process.
To explore this issue, we will first generate artificial data
using the Monte Carlo (MC) method. The von Neumann rejection method
is employed to select the random data points that follow the normalized
distribution of the $\pi^0\pi^0$ in range between 270~MeV and
290~MeV predicted in the NREFT. The MC data are generated using the
central value of $G_0/G_1$ given in Eq.~\eqref{eq:G01ratio}, and
$a_0-a_2=0.2640$ and $a_2=-0.0444$~\cite{Gasser:2009zz} in units of
$M_{\pi^+}^{-1}$ as input. These data can then be divided into a
number of bins with the statistical errors given by the square root
of the number of events in each bin. Varying the MC event numbers
and the bin widths, one may investigate the impact of the event
numbers as well as the experimental energy resolution on the
precision of the extraction of the scattering lengths.

\begin{figure}[t]
\centering
\includegraphics[width=\linewidth]{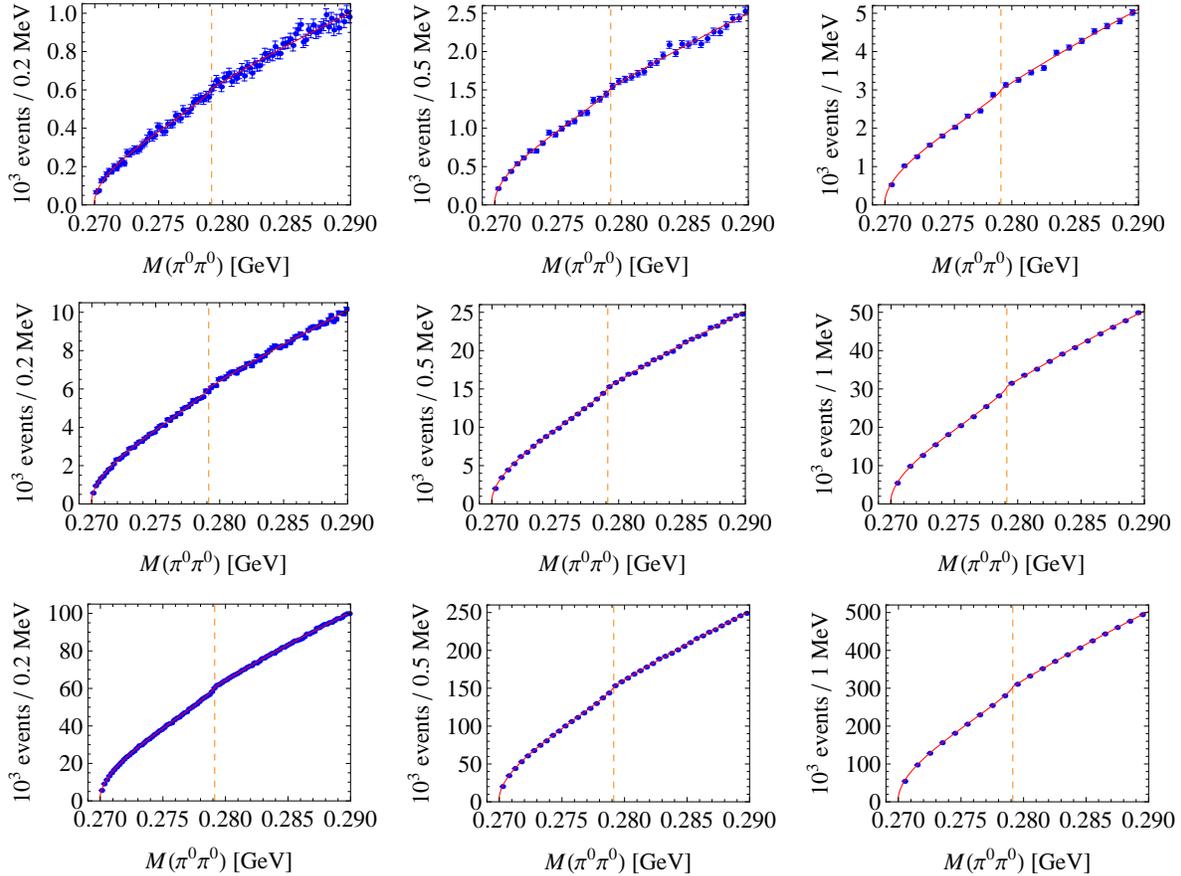}
\caption{Various sets of MC events and their best fits. The event
numbers in the range of $[270,290]$~MeV of the first, the second and
the third rows are about $6\times10^4$, $6\times10^5$ and
$6\times10^6$, respectively. The bin widths of the first, the second
and the third columns are 0.2, 0.5 and 1~MeV, respectively. The
vertical dashed line indicates the $\pi^+\pi^-$
threshold.}\label{fig:fit}
\end{figure}

\begin{table}[t]
\begin{center}
   \begin{tabular}{|l | c | c c c c |} \hline
      Bin width & {Events} & $6\times10^4$       & $6\times10^5$ & $3\times10^6$ & $6\times10^6$     \\
      \hline
      \multirow{2}{*}{0.1~MeV} & $\chi^2/dof$     & 1.21    & 1.09  & 1.16  & 0.88  \\\cline{2-6}
       & $a_0-a_2$ & $0.293\pm0.036$ & $0.260\pm0.012$ & $0.2717\pm0.0048$ & $0.2661\pm0.0036$  \\
       \hline
      \multirow{2}{*}{0.2~MeV} & $\chi^2/dof$     & 0.72    & 1.15  & 1.05  & 1.12  \\\cline{2-6}
       & $a_0-a_2$ & $0.286\pm0.035$ & $0.251\pm0.014$ & $0.2722\pm0.0048$ & $0.2621\pm0.0038$  \\
       \hline
      \multirow{2}{*}{0.5~MeV} & $\chi^2/dof$     & 0.93    & 0.54  & 1.27  & 1.30  \\\cline{2-6}
       & $a_0-a_2$ & $0.262\pm0.026$ & $0.256\pm0.012$ & $0.2659\pm0.0051$ & $0.2693\pm0.0035$  \\
       \hline
      \multirow{2}{*}{1~MeV} & $\chi^2/dof$     & 1.05    & 0.78  & 1.17  & 0.69  \\\cline{2-6}
       & $a_0-a_2$ & $0.221\pm0.054$ & $0.291\pm0.010$ & $0.2658\pm0.0054$ & $0.2661\pm0.0037$  \\
       \hline
      \multirow{2}{*}{2~MeV} & $\chi^2/dof$     & 0.59    & 1.06  & 1.05  & 1.37  \\\cline{2-6}
       & $a_0-a_2$ & $0.260\pm0.040$ & $0.262\pm0.012$ & $0.2592\pm0.0055$ & $0.2632\pm0.0037$  \\ 
       \hline
   \end{tabular}
   \caption{\label{tab:fit} Results of fitting to various sets of MC data. The extracted scattering lengths are given
   in units of $M_{\pi^+}^{-1}$.}
\end{center}
\end{table}

We tried a number of different combinations of the event numbers and bin widths.
Figure~\ref{fig:fit} shows the ones with about $6\times10^4$, $6\times10^5$ and
$6\times10^6$ events in the range of $[270,290]$~MeV. We then fit the $\pi^0\pi^0$
invariant mass distribution calculated using Eq.~\eqref{eq:NRT} to the MC data.
$G_0/G_1$ is fixed to the same value used in the data generation.
The free
parameters are an overall normalization constant, $a_0-a_2$ and $a_2$. The results
of the fits are collected in Table~\ref{tab:fit}, where the uncertainties only
reflect the statistical errors in the fit. Because of the random fluctuation in the
data generating process, the best fit values are not guaranteed to be the same as
the input. An interesting observation is that the precision of the extraction seems
to be quite insensitive to the bin widths, at least up to
2~MeV. Comparing the extracted values with the
input $M_{\pi^+}(a_0-a_2)=0.2640$ and $M_{\pi^+}a_2=-0.0444$, one sees that the
precision of the extracted value of $a_0-a_2$ can reach $10-20\%$ for $6\times10^4$ events
in the range of $[270,290]$~MeV. For more events, the precision is better by  a
factor of around $\sqrt{\mathcal{N}/\mathcal{N}'}$, with $\mathcal{N}'$ and
$\mathcal{N}$ the new and old event numbers, as it should be. From
Table~\ref{tab:fit}, one sees that the statistical precision of
Ref.~\cite{Batley:2000zz}, $M_{\pi^+}(a_0-a_2)=0.2571\pm0.0048(\text{stat.})$, may
be reached with $3\times10^6$ events. The spectrum is rather insensitive to
$a_2$ such that the uncertainty is about 50\% for $6\times10^6$ events.
In fact, because $a_2$, independent of $a_0-a_2$, only enters through the $\pi^0\pi^0\to\pi^0\pi^0$
rescattering, its effect can be largely absorbed into the polynomial production amplitudes.
We have checked that if $G_0/G_1$ was released as an additional free parameter, one would not get any useful
information on $a_2$ any more. Furthermore, the uncertainty of $a_0-a_2$ would also increase
by a factor of about 2 to 2.5.
In fact, $G_0/G_1$ also contributes to the $\Upsilon(3S)\to\Upsilon(2S)\pi^+\pi^-$,
and its value can be extracted by measuring the $\pi^+\pi^-$ spectrum in parallel,
and thus not completely free.
Notice that if we use $a_0$ instead of $a_0-a_2$ as a free
parameter, the errors would be much larger. We should stress that to the precision at
per cent level, radiative corrections, which is most important in the $\pi^+\pi^-$ threshold
region, should be taken into account~\cite{Bissegger:2008ff,Kubis:2009sb}.
Nevertheless, we expect that the precision that can be achieved would not get worsened as the
photon-exchange can be taken into account by a simple replacement of the $\pi^+\pi^-$
loop function~\cite{Kubis:2009sb}. Since our aim
is to explore the possibility of extracting $a_0-a_2$, they will not be considered here.

From the CLEO data of the $\Ups$, the
events in the range of $[270,290]$~MeV correspond to around 15\% of the total yield of the
process. Thus, $6\times10^4$, $6\times10^5$, $3\times10^6$ and $6\times10^6$ events
in $[270,290]$~MeV require the yields of the $\Ups$ of about $4\times10^5$,
$4\times10^6$, $2\times10^7$ and $4\times10^7$, respectively. Given that the
branching fraction of  $\Ups$ is $(1.85\pm0.14)\%$~\cite{PDG}, at least 2
billion $\Upsilon(3S)$ events have to be accumulated in order to obtain $6\times10^6$ events
in the range of $[270,290]$~MeV. Other factors like the detecting efficiency will increase the
number even further.

\section{Summary}
\label{sec:sum}

In this paper, we investigate the possibility of extracting the $\pi\pi$ $S$-wave
scattering lengths using the cusp effect in heavy quarkonium transitions emitting
two neutral pions. These processes are different from all the others for the cusp
effect because all involved particles apart from the pions are
heavy. This has a theoretical advantage that the cross-channel rescattering of a pion off a heavy quarkonium
is weak due to the OZI suppression. Due to the approximate isospin symmetry,
$\mathcal{B}(V'\to V \pi^+\pi^-)/\mathcal{B}(V'\to V \pi^0\pi^0) \sim 2$ will lead to
an enhanced cusp effect in $\pi^0\pi^0$ invariant mass spectrum. Since we are dealing
with the process where the relevant particles have low momenta, the
framework of NREFT is
adopted to calculate the decay amplitude which can be directly parameterized in terms of the
$\pi\pi$ threshold parameters. In the present analysis, we worked out
the amplitude to order
$\mathcal{O}(a_{\pi\pi}\epsilon^2)$.

We then focus on the $\Ups$, for which the parameters in the production amplitude
are determined by matching to a fit to the experimental data based on the
chiral unitary approach. In order to have a feeling on the achievable
accuracy of the extraction of the scattering length, we generated a
number of sets of artificial data
using the Monte Carlo method. We then fitted these synthetic data to
using the values of $a_0-a_2$
and $a_2$ as free parameters. It is comforting  to see that the
resulting accuracy is insensitive
to the bin width and energy resolution. A statistical precision of about 2\% and
1.5\% of $a_0-a_2$ can be reached with $2\times10^7$ and $4\times10^7$ events of
the $\Ups$, which corresponds to at least $1\times10^9$ and $2\times10^9$
$\Upsilon(3S)$ events, respectively. The precision can be worsened by a factor of about 2
in reality because $G_0/G_1$ in the production amplitude cannot be fixed completely.
However, measuring the $\Upsilon(3S)\to \Upsilon(2S)\pi^+\pi^-$ in parallel is very
helpful in constraining $G_0/G_1$, and hence increasing the precision of the $a_0-a_2$ extraction.
The CLEO detector already recorded a sample of
$(5.93\pm0.10)\times10^6$ $\Upsilon(3S)$ decays~\cite{Bhari:2008aa},
while this number
is $1.08\times10^8$ for the BaBar detector~\cite{Lees:2011bv}. With future
high-luminosity $B$-factories, the sample can be one or two order-of-magnitude
larger.

\subsection*{Acknowledgments}

We would like to thank Bastian Kubis for very useful discussions and a careful reading,
and Ulf-G. Mei{\ss}ner for comments on the manuscript.
This work is supported in part by the EU HadronPhysics3 project ``Study
of strongly interacting matter'', by the European Research Council (ERC-2010-StG
259218 NuclearEFT), by the DFG and the NSFC through funds provided to the
Sino-German CRC 110 ``Symmetries and the Emergence of Structure in QCD'' and by the
NSFC (Grant No. 11165005).

\begin{appendix}

\section{An estimate of the $J/\psi \pi$ scattering length}
\label{sec:app}

\begin{figure}[t]
\centering
\includegraphics[width=0.7\textwidth]{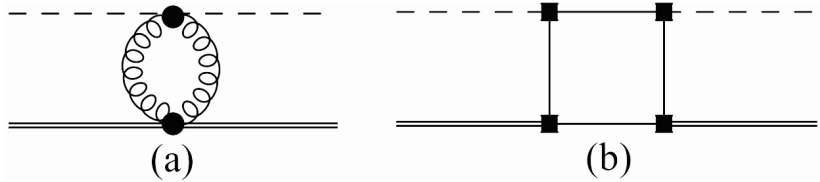}
\caption{Schematic diagrams of the charmonium-pion scattering. Here the doubly-solid, dashed,
solid and wiggly lines
represent charmonia, pions, charmed mesons and gluons, respectively.}
\label{fig:psipi}
\end{figure}
Before presenting the formalism, let us first roughly estimate the $J/\psi\pi$
scattering length. Certainly there are no scattering data available, but one
may use the amplitude of $\psi'\pi\to J/\psi\pi$ as a reference for the $J/\psi\pi$
elastic scattering amplitude. Considering a process of scattering a pion off a
charmonium, two possible mechanisms are shown in Fig.~\ref{fig:psipi}: (a)
corresponds to the situation in which the charmonium emits two soft gluons which hadronize into
pions. This mechanism can be described by using the method of QCD multipole
expansion. The charmonium-pion scattering can also occur through intermediate
charmed mesons, as depicted in Fig.~\ref{fig:psipi}(b), which represents a kind of
non-multipole effect~\cite{Zhou:1990ik}. Noticing that the analytic structures of
the amplitudes for these two mechanisms are different, one concludes that there is no double counting.

In the first mechanism, the difference between the transition ($\psi'\pi\to
J/\psi\pi$) and elastic ($J/\psi\pi\to J/\psi\pi$) amplitudes is due to the
charmonia-two-gluon vertex, which is proportional to a quantity called charmonium
chromo-polarizability $\alpha_{c\bar c}$, the definition of which can be found in
Ref.~\cite{Sibirtsev:2005ex}. Because $\alpha_{\psi'} \alpha_{J/\psi} \geq
|\alpha_{\psi'J/\psi}|^2$~\cite{Sibirtsev:2005ex}, one may expect that the elastic
amplitude is somewhat larger than the transition one, i.e.,
$|\mathcal{A}(J/\psi\pi\to J/\psi\pi)_{(a)}| \gtrsim |\mathcal{A}(\psi'\pi\to
J/\psi\pi)_{(a)}|$ at the $J/\psi\pi$ threshold. In the second mechanism, the
elastic $J/\psi\pi$ scattering amplitude is proportional to $g_2^2$, and the
transition amplitude is proportional to $g_2 g_2'$, where $g_2(g_2')$ are the
$J/\psi(\psi') D\bar D$ coupling constants. Neither of these coupling constants can
be measured directly. Based on a vector dominance model, it was estimated in
Ref.~\cite{Colangelo:2003sa} that $g_2=\sqrt{M_{J}}/(M_D f_{J/\psi})$ with
$f_{J/\psi}$ the $J/\psi$ decay constant. Thus, one may estimate
$$\frac{g_2'}{g_2}\approx \frac{f_{J/\psi}}{f_{\psi'}} \approx \left( \frac{\Gamma(J/\psi\to e^+e^-)}
{\Gamma(\psi'\to e^+e^-)}
\right)^{1/2} \approx 1.5,
$$
which means $|\mathcal{A}(\psi'\pi\to J/\psi\pi)_{(b)}| \gtrsim
|\mathcal{A}(J/\psi\pi\to J/\psi\pi)_{(b)}|$. Combining with the estimate for the
first mechanism, it is reasonable to assume
\begin{equation}
    |\mathcal{A}(J/\psi\pi\to J/\psi\pi)| \sim |\mathcal{A}(\psi'\pi\to J/\psi\pi)|
    \label{eq:AJpsipi}
\end{equation}
at the $J/\psi\pi$ threshold.

\begin{figure}[t]
\centering
\includegraphics[width=0.6\textwidth]{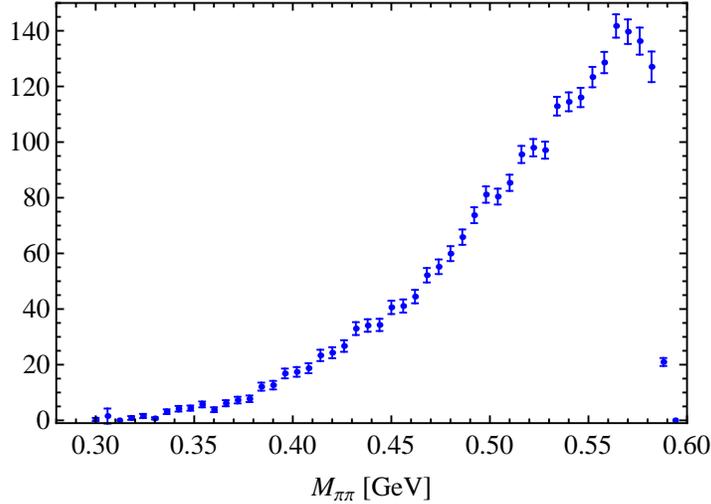}
\caption{Phase space subtracted invariant mass spectrum of the $\pi\pi$ system
for the decay $\psi'\to J/\psi\pi^+\pi^-$ (in arbitrary units). The original data are
taken from Ref.~\cite{Bai:1999mj}.}
\label{fig:psisub}
\end{figure}
Because of crossing symmetry, the $\psi'\pi\to J/\psi\pi$
scattering amplitude is related to the $\psi'\to J/\psi\pi\pi$.
Assuming that the amplitudes are constant, denoted by
$\widetilde{C}$, they are the same for scattering and decay
processes. This assumption is definitely not realistic, but it can
be used to place an upper limit of the $J/\psi\pi$ scattering
length. The $J/\psi\pi$ threshold occurs when the $\pi\pi$ invariant
mass is $\sqrt{s_3}=M_{\pi\pi}=415$~MeV. In Fig.~\ref{fig:psisub},
we show the phase-space-subtracted invariant mass spectrum of the
$\pi\pi$ system for the decay $\psi'\to J/\psi\pi^+\pi^-$. That is,
the experimental data~\cite{Bai:1999mj} are divided by
$|\vec{p}_1^{\,*}||\vec{p}_3|$, where
\begin{eqnarray}
    |\vec{p}_1^{\,*}| = \frac1{2\sqrt{s_3}} \sqrt{\lambda\left(s_3,M_{J}^2,M_{\pi}^2\right)}, \qquad
    |\vec{p}_3| = \frac1{2M_{\psi'}} \sqrt{\lambda\left(M_{\psi'}^2,s_3,M_{J}^2\right)}.
\end{eqnarray}
From Fig.~\ref{fig:psisub}, one can see that the physical decay
amplitude at $\sqrt{s_3}=415$~MeV should be smaller than the assumed
(nonrealistic) constant amplitude $|\widetilde{C}|$. From the decay
width of $\psi'\to J/\psi\pi^+\pi^-$, one can extract the constant
$|\widetilde{C}|\approx9.6$. Using Eq.~\eqref{eq:AJpsipi}, we get an
approximate upper limit for the $J/\psi\pi$ $S$-wave scattering
length
\begin{equation}
    \left|a_{J/\psi\pi} \right| \lesssim \frac{|\widetilde{C}|}{8\pi \left( M_J+M_\pi \right)} \approx 0.02~\text{fm}.
\end{equation}
Similarly, using the measured decay width of the
$\Upsilon(3S)\to\Upsilon(2S)\pi^+\pi^-$, we get
\begin{equation}
    \left|a_{\Upsilon(2S)\pi}\right| \lesssim 0.01~\text{fm}.
\end{equation}

\end{appendix}

\end{document}